\documentclass[a4paper,11pt]{article}
\usepackage{jinstpub} 
\pdfoutput=1

\usepackage{graphicx}

\usepackage{natbib}

\title{\boldmath Performance testing of a novel short axis photomultiplier tube for the HUNT project}

\author[a,b]{Yijiang Peng,}
\author[c,e]{Zike Wang,}
\author[c,d]{Bo Gao,}
\author[f]{Yiyue Tang,}
\author[c,d]{Mingjun Chen,}
\author[c,d]{Kai Li,}
\author[g]{Ling Ren,}

\author[c,d,1]{Xiaohao You\note{Corresponding author.}}
\author[a,b,1]{and Maoyuan Liu}

\affiliation[a]{College of Science, Tibet University,\newline Lhasa 850000, China}

\affiliation[b]{Key Laboratory of Cosmic Rays (Tibet University), Ministry of Education,\newline Lhasa 850000, China}

\affiliation[c]{Key Laboratory of Particle Astrophysics, Institute of High Energy Physics, Chinese Academy of Sciences,\newline Yuquan Rd, Beijing, China}

\affiliation[d]{Tianfu Cosmic Ray Research Center, Institute of HighEnergy Physics,\newline Chengdu, Sichuan, China}

\affiliation[e]{School of Physical Sciences, University of Chinese Academy of Sciences,\newline Yanqihu East Rd, Beijing, China}

\affiliation[f]{School of Physical Science and Technology, Southwest Jiaotong University,\newline Chengdu, 611756, Sichuan Province, China}

\affiliation[g]{North Night Vision Science \& Technology (Nanjing) Research Institute Co. Ltd,\newline Kangping St, Nanjing, China}

\emailAdd{youxiaohao@ihep.ac.cn}
\emailAdd{liumy@utibet.edu.cn}

\abstract{Photomultiplier tubes (PMTs) with large-area cathodes are increasingly being used in cosmic-ray experiments to enhance detection efficiency. The optical modules (OMs) of the High-Energy Underwater Neutrino Telescope (HUNT) have employed a brand new N6205 20-inch microchannel plate photomultiplier tube (MCP-PMT) developed by the North Night Vision Science \& Technology (Nanjing) Research Institute Co. Ltd. (NNVT). In order to make the 20-inch PMT fit into the 23-inch diameter pressure-resistant glass sphere, NNVT improved the internal structure of PMT and shortened the height of PMT by more than 10~cm. The first batch of these PMTs has been delivered for preliminary research work. This paper describes a specific PMT testing platform built for the first batch of 15 MCP-PMTs, and some performance parameters of PMT, such as peak-to-valley ratio, TTS and nonliniearity, are measured. The measurement results show that the new PMT still has good performance and can meet the requirements of HUNT project.}

\keywords{Photon detectors for UV, visible and IR photons (vacuum); Neutrino detector; Large detector systems for particle and astroparticle physics, Cherenkov detectors}

\begin{document}
\maketitle

\flushbottom

\section{Introduction}
\label{sec:1}

Since Moisey Markov in 1960 proposed deploying large-volume detectors in deep lakes or seas to detect Cherenkov radiation generated by high-energy neutrinos \cite{1960}, many large Cherenkov high-energy neutrino telescope experiments have emerged internationally, such as DUMAND \cite{DUMAND}, AMANDA \cite{AMANDA}, ANTARES \cite{ANTARES}, IceCube \cite{IceCube}, and Baikal-GVD \cite{Baikal-GVD}. The High-Energy Underwater Neutrino Telescope (HUNT) is a next-generation high-energy neutrino detector proposed by the Institute of High Energy Physics (IHEP) of the Chinese Academy of Sciences. Its primary goal is to observe high-energy neutrinos in association with the discovery of a large number of ultra-high-energy gamma photons in the Milky Way by LHAASO \cite{lhaaso-nature}, thereby helping to solve the mystery of the origin of cosmic rays. The HUNT project plans to deploy over 55,000 Optical Modules (OMs) in an array within a volume of 30 $km^3$, with each OM housing a 20-inch PMT.

The PMT is the most commonly used photosensitive detector, and can provide the function of detecting Cherenkov light for OMs of high-energy neutrino telescopes. In order to ensure the highest possible photon detection efficiency, the photosensitive device must have the largest possible collection area and the highest possible optical sensitivity. In order to achieve good angular resolution, it should have good time accuracy. Utilizing a PMT with unprecedented large photocathode area for the detection of high-energy neutrinos provides HUNT with high sensitivity, low dark counting rate, and wide dynamic range, which has advantages, especially in large-area light-signal detection and position resolution. The performance of a single PMT will directly affect the detection capability of the entire array. Considering the PMT manufacturing process and based on the physical requirements of HUNT, the PMT needs to meet the performance indicators specified in table~\ref{tab:pmt parameter}.

Each PMT will undergo at least one performance test after delivery, including single photoelectron (SPE) charge spectrum (gain, SPE resolution, and peak-to-valley ratio), transit time spread (TTS), dark noise rate (DNR), Q-T effect curve, anode charge nonlinearity, and after pulse ratio. These test results will be used not only to evaluate the quality of the PMT itself, but also to provide data support for optimizing the HUNT detection array. This paper introduces the basic information about N6205, describes the construction of the testing platform, and discusses in detail the testing process and calculated results for each feature.

\begin{table}[ht]
\centering
\caption{Performance requirements for 20-inch PMT in HUNT}
\begin{tabular}{l l l} 
  \hline
  Parameter & Specification \\ 
  \hline
  Gain & 5 $\times$ 10$^6$ \\ 
  High voltage & $<$2000 V, Mean $\pm$ 120 V \\ 
  Peak-to-valley & $>$2 \\ 
  QE & $>$25\% \\ 
  Rise time & $<$3 ns, typically 1.4 ns \\ 
  TTS & $<$7 ns \\ 
  DNR & $<$25 kHz, threshold 1/3 PE \\ 
  Afterpulse ratio & $<$2\%, 0.5 $\mu$s - 10 $\mu$s, typically 1\% \\ 
  Anode charge non-linearity & $<$1800 PEs (10\%), mean $>$2000 PEs \\ 
  Stability & variation gain $<$5\% \& DNR $<$10\%, 3-12 h \\ 
  \hline
\end{tabular}
\label{tab:pmt parameter}
\end{table}

\section{N6205 MCP-PMT}
\label{sec:2}

The pressure vessel designed for HUNT is a spheroidal glass chamber with a maximum diameter of 23 inches (585 mm). Therefore, based on NNVT's previously developed 20-inch PMT \cite{NNVT}, IHEP and NNVT jointly designed a new 20-inch MCP-PMT (N6205) with an axial height of 470 mm, allowing it to be installed in the glass chamber. N6205 uses a dual alkali photocathode with high quantum efficiency. The unique lotus shaped expansion type focusing electrode greatly improves the efficiency of photoelectrons reaching the microchannel plate. PMT uses a negative high-voltage circuit ,connected via a 50 $\Omega$ coaxial cable simultaneously transmits high voltage and signals. It has passed temperature aging tests. Under the gain condition of 5 $\times$ 10$^6$, the working voltage is less than 2000 V. 

\begin{figure}[ht]
\centering
\includegraphics[width=.35\textwidth]{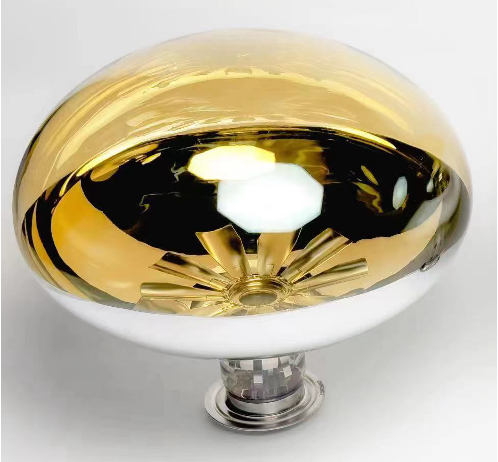}
\qquad
\includegraphics[width=.35\textwidth]{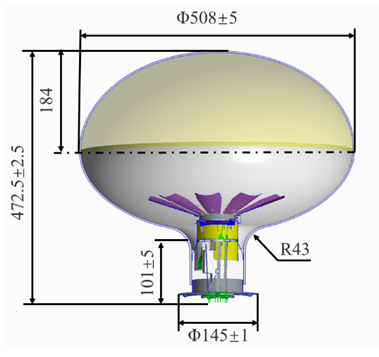}
\caption{Left: the appearance of N6205. Right: the specific dimensions in mm.}
\label{fig:pmt}
\end{figure}

\begin{figure}[ht]
    \centering
    \includegraphics[width=0.8\linewidth]{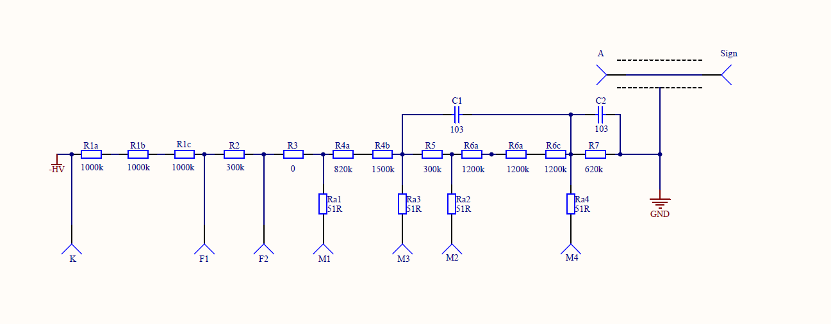}
    \caption{Circuit design diagram of the voltage divider.}
    \label{fig:base}
\end{figure}

The specific dimensions and appearance of N6250 are shown in figure~\ref{fig:pmt}. The circuit design of the voltage divider used in N6205 is shown in figure~\ref{fig:base}; it has a total resistance of 10 M$\Omega$.

\section{Platform}
\label{sec:3}

\begin{figure}[ht]
    \centering
    \includegraphics[width=0.6\linewidth]{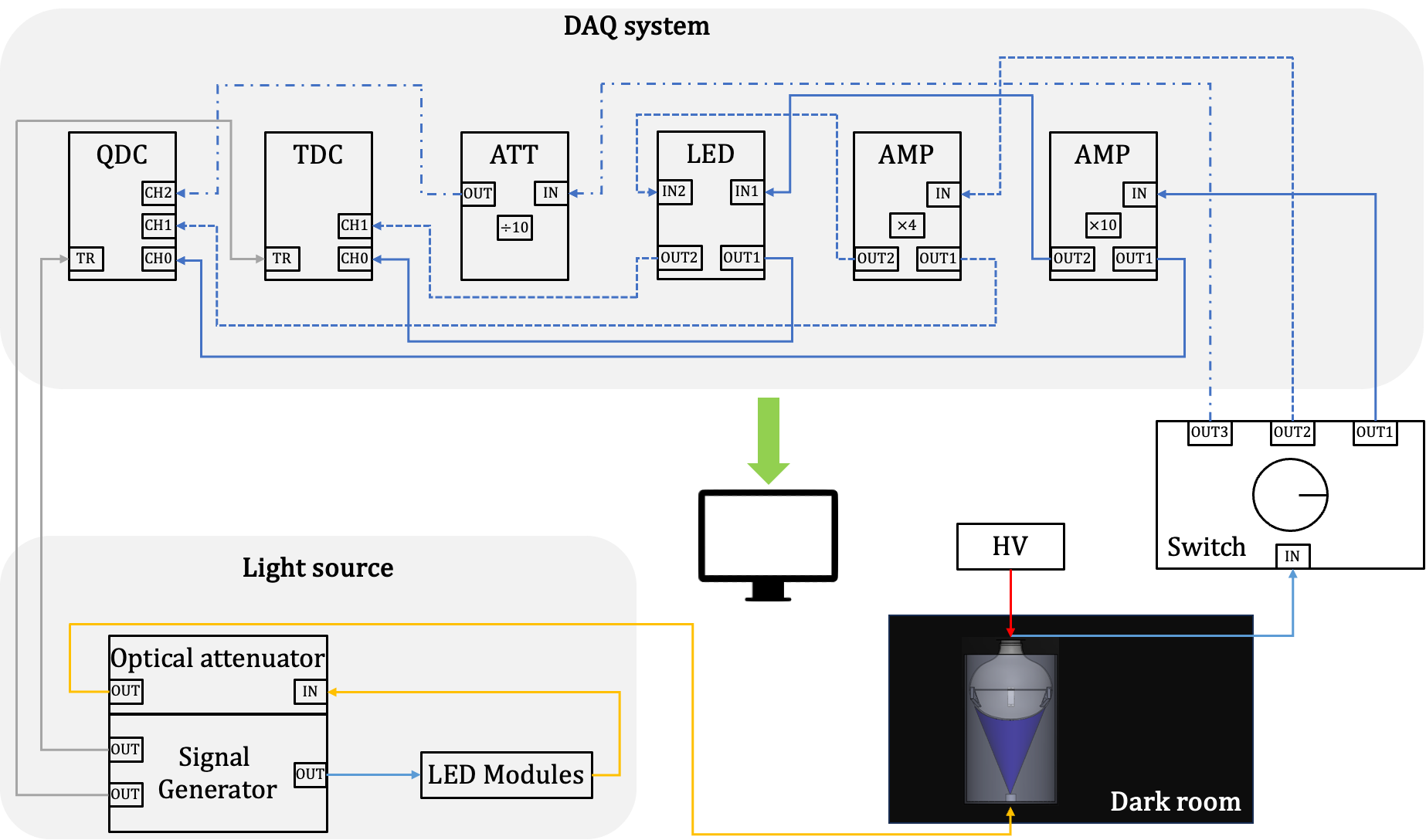}
    \caption{Schematic diagram of testing platform. The yellow line represents the optical fiber, the red line represents the HV line, the gray line represents the trigger signal, and the blue line represents the signal line. Different types of blue lines represent different test proseses.}
    \label{fig:daq}
\end{figure}

The testing platform, as shown in figure~\ref{fig:daq}, consists of three parts: darkroom, light source, and DAQ system. This section provides a detailed introduction  to the three parts.

\subsection{Dark room}
\label{subsec:3.1 dark room}

\begin{figure}[ht]
    \centering
    \includegraphics[width=0.25\linewidth]{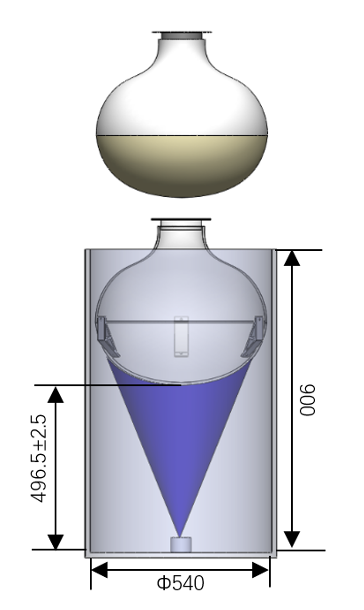}
    \caption{Dimensions of the barrel placed in the darkroom to support PMT testing.}
    \label{fig:barrel}
\end{figure}

In the laboratory, a separate room was set up as a darkroom, where two barrels were installed for testing the PMTs. The walls of the darkroom and the outer wall of each barrel were pasted with iron-based amorphous alloy foils \cite{shied} for shielding the magnetic field, resulting in a residual magnetic field at the PMT cathode position of less than 5 $\mu$T. The magnetic shielding material is black, and a black cloth is pasted inside each barrel to eliminate the influence of stray light. The size of a barrel is shown in figure~\ref{fig:barrel}. It has an inner diameter of 540 mm and a height of 900 mm. Three 135° brackets were evenly installed on the inner wall of each barrel at a height of 670 mm for PMT placement. For these setting, the top of the PMT cathode is 500 mm above the base, and the PMT equator is 670 mm above the base.

\subsection{Light source}
\label{subsec:3.2 Light source}

The testing system uses a continuous optical attenuator integrated with a multi-channel signal generator. This system can not only output light pulse signals, it can also allow continuous adjustment of light power by three orders of magnitude through software control \cite{OPatt1,OPatt2}.

\begin{figure}[ht]
\centering
\includegraphics[width=.45\textwidth]{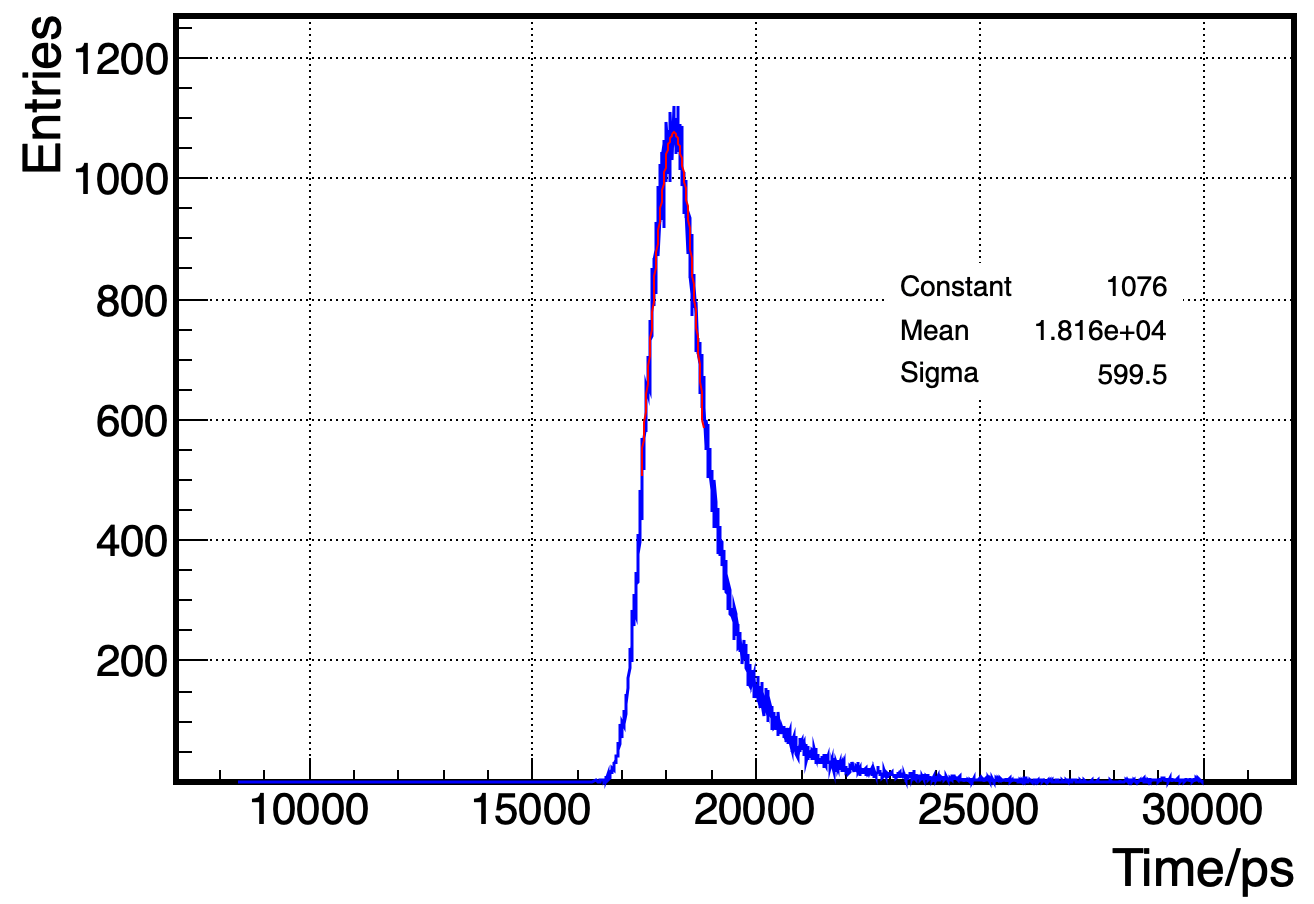}
\qquad
\includegraphics[width=.45\textwidth]{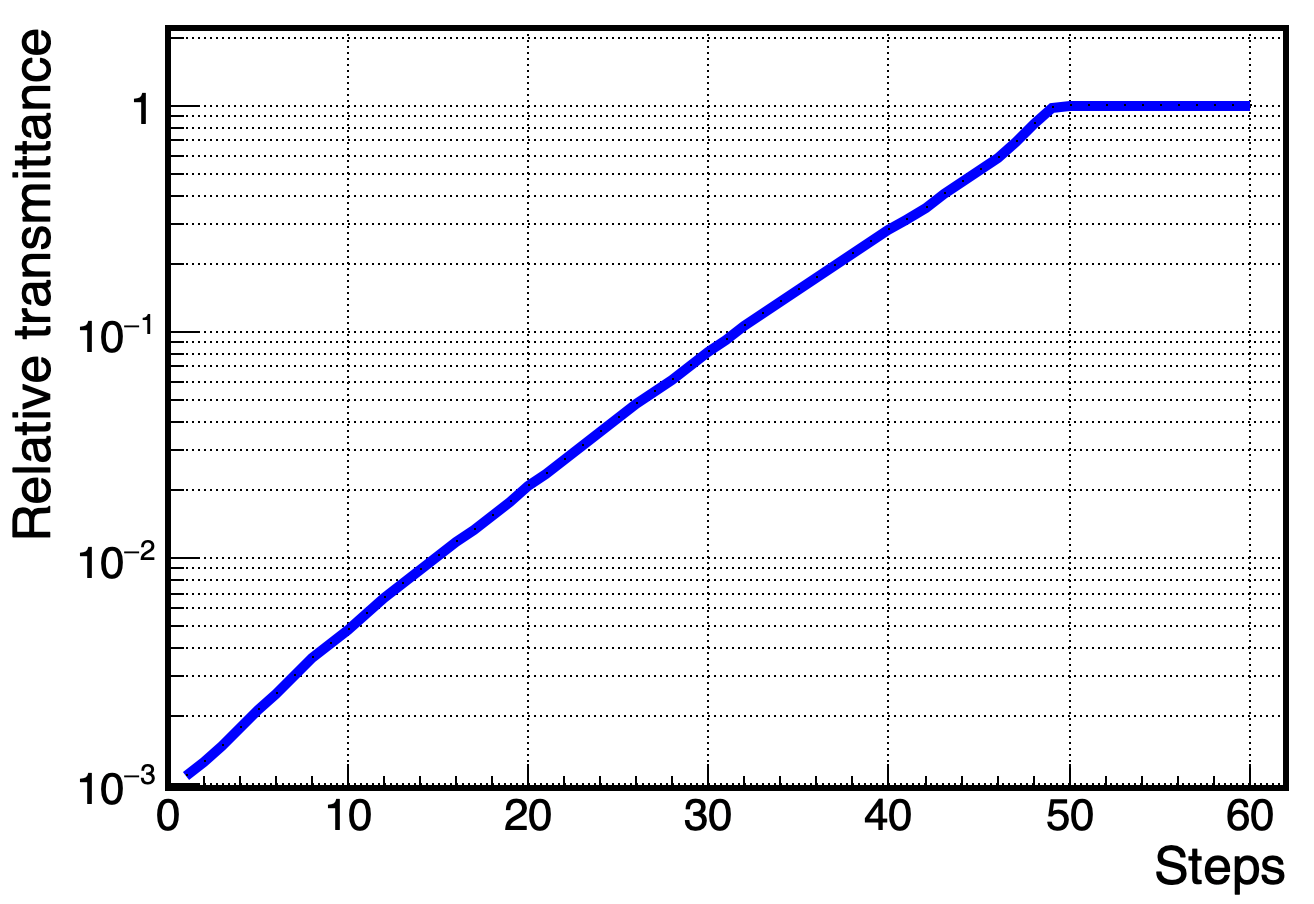}
\caption{Left: The emission time of the 405-nm LED used for testing. Right: The relationship between step size and light relative transmittance when the motor rotates one revolution.}
\label{fig:led time and light att}
\end{figure}

The signal generator features 10 channels with SMA connectors, each offering independent adjustment of pulse width and delay. The output signal has an amplitude range of 0 to 6.6 V, and after calibration, the relative timing jitter between channels is maintained below 50 ps. The triggering signal is input to a 405 nm LED module to drive its emission; the calibrated emission time is shown in figure~\ref{fig:led time and light att}. The optical signal is then input to a continuous optical attenuator via a quartz optical fiber.

The core components of the continuous optical attenuator consist of a neutral-density continuous attenuation plate and a deceleration stepper motor that drives the rotating attenuation plate. The base material of the attenuation plate is custom-made quartz glass with a pinhole diameter  of 10 mm, an outer diameter of 100 mm, and a thickness of 1.0 mm. A semi-transparent film with a variable thickness is coated on the surface of the glass to achieve continuous variation of light transmittance with angle. The attenuation plate is sealed in a black aluminum dark chamber, with the pinhole fixed on the shaft of the stepper motor. On both sides of the dark chamber, quartz lenses with SMA-905 interfaces are fixed; the optical axes of the two lenses coincide. After connecting the lenses to the quartz optical fiber, the relative light transmittance is calibrated with a 0.1\%-accuracy optical power meter under room temperature conditions, and the relationship between the relative light transmittance and the number of motor rotation steps is obtained. As shown in figure~\ref{fig:led time and light att}, these data provide assistance for Q-T effect testing and anode charge nonlinearity testing.

The light source device serves three principal functions: Channels 1 and 2 act as trigger windows for QDC and TDC to generate signals; Channels 3-9 can all serve as trigger signals for the LED module; and the optical attenuator continuously attenuates the optical signal, enabling adjustable control of light intensity.

The output light is split into two parts by a UV-transmitting quartz optical fiber, and each part is input into two beam expanders. A fiber beam expander is installed at the central position inside the barrel to expand the output light, forming a spot with a diameter of approximately 450 mm on the cathode surface of the PMT, covering over 80\% of its area.

\subsection{DAQ system}
\label{subsec:3.3 daq system}

Figure~\ref{fig:daq} shows the setup method of the DAQ system. A divider distributes HV to the anode pin of the PMT and receives pulse signals from the anode, which are then transmitted to a custom-made selector switch. Each channel of the switch has one input and three outputs, with the output signals connected to different modules in the VME crate for data processing in various test projects. The detailed procedure is explained in Section~\ref{sec:4}. All tests were conducted using commercial instruments, with the various models and parameters listed in Table~\ref{tab:DAQs}. The processed data are transmitted to the control computer for storage and analysis via a USB 3.0 communication cable.

\begin{table}[ht]
\centering
\caption{Devices and specific performance used during testing (1 fC = 1 $\times$ 10$^{-15}$, LSB means Least Significant Bit).}
\smallskip
\begin{tabular}{l l l} 
  \hline
  Instrument & Model & Characteristics \\ 
  \hline
  Oscilloscopes & PICOSCOPE 6824E & 8  Channels, 500 MHz bandwidth \\ 
  Power Supply & CAEN DT8033M & 8 Channels, 4 kV/3 mA (6 W) \\ 
  QDC & CAEN V792N & 16 Channels, 100 fC/LSB \\ 
  TDC & CAEN V775N & 16 Channels, 35-300 ps/LSB \\ 
  TDC & CIQTEK TDC0110 & 2 Channels, 8 ps/LSB \\ 
  Attenuator & CAEN V859 & 0-89 dB attenuator \\ 
  LED & CAEN N841 & -1 to -255 mV (1 mV step) \\ 
  Amplifier & CAEN N978 & 1-10 variable gain \\ 
  Amplifier & CAEN N979 & 10 fixed gain \\ 
  \hline
\end{tabular}
\label{tab:DAQs}
\end{table}

\section{Test process}
\label{sec:4}

The entire testing process is divided into seven following stages: waveform inspection, cooling, SPE charge spectrum, TTS, Q-T effect, anode charge nonlinearity, and afterpulse ratio.

\subsection{Waveform inspection}
\label{subsec:4.1 waveform check}

\begin{figure}[h]
    \centering
    \includegraphics[width=0.45\linewidth]{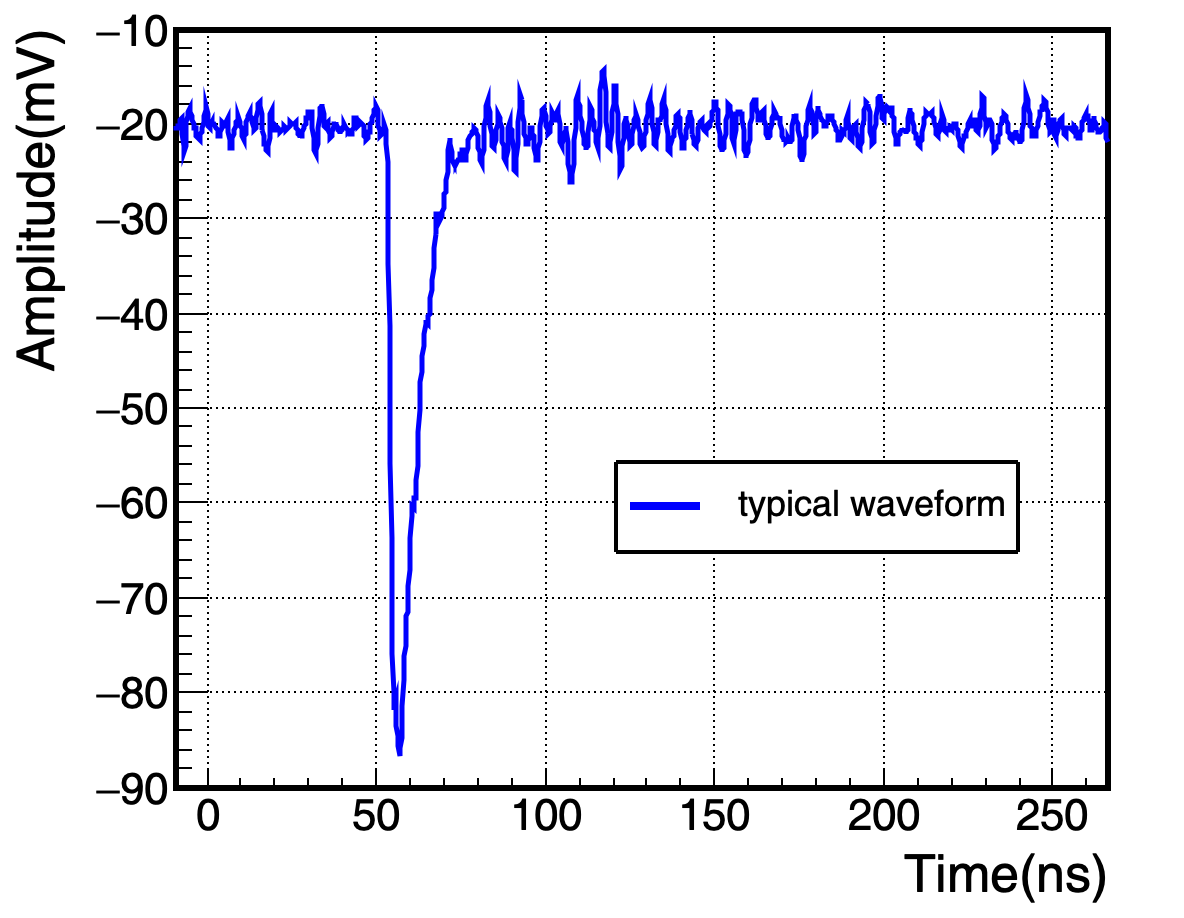}
    \caption{The waveform under SPE conditions after being amplified by a factor of ten, showing an amplitude of 60 mV.}
    \label{fig:waveform}
\end{figure}

To inspect the pulse signal output by the PMT, the light intensity was adjusted to a level of SPE, with a signal trigger frequency of 1 kHz. The PMT signal was acquired using an oscilloscope, PICOSCOPE 6824E, with results shown in figure~\ref{fig:waveform}. The blue waveform is the PMT signal. When checking the signal, it was found that there was overshoot in the PMT signal. Since directly connecting to the QDC would cause instrument damage, the baseline was biased to -2 mV, keeping the PMT signal below 0 mV.

\subsection{Cooling}
\label{subsec:4.2 system stability}

Large-sized PMT usually needs to be cooled for several hours before testing to stabilize the dark noise, so as to improve the signal-to-noise ratio and stability during testing~\cite{coolingreason}. When cooling tests are initially conducted, it is essential to ensure that the tests are performed under SPE conditions. Therefore, before running the test script, a debugging script is executed to determine the HV and trigger ratio during testing. The trigger ratio is defined as the number of events in the SPE charge spectrum charge larger than $\mu_{ped}$ + 5$\sigma$ to the total events. $\mu_{ped}$ and $\sigma$ are the mean and standard deviation obtained by Gaussian fitting of pedestal. When the trigger ratio is 10\%, according to Poisson statistics, this ensures that the recorded signal contains over 95\% SPE. During debugging, the high voltage is set initially according to the values provided in the NNVT factory test table, which calls for taking a 500,000 trigger signal, and repeating four times. The output includes the statistical average and error of the trigger ratio, gain, and charge. The high voltage and other settings are adjusted to achieve trigger ratios and gains at levels of 10\% and 5×10$^6$, respectively, before running the formal test script.

The cooling test includes monitoring four parameters: trigger ratio, gain, charge, and DNR. This serves as an evaluation data set and is conducted over a period of at least 3 hours. The signal output from the PMT is routed through a switch, with the switch set to the first gear. The combined signal is then amplified tenfold using an N979 amplifier before being processed by the V792N.

\begin{figure}[h]
    \centering
    \includegraphics[width=0.48 \linewidth]{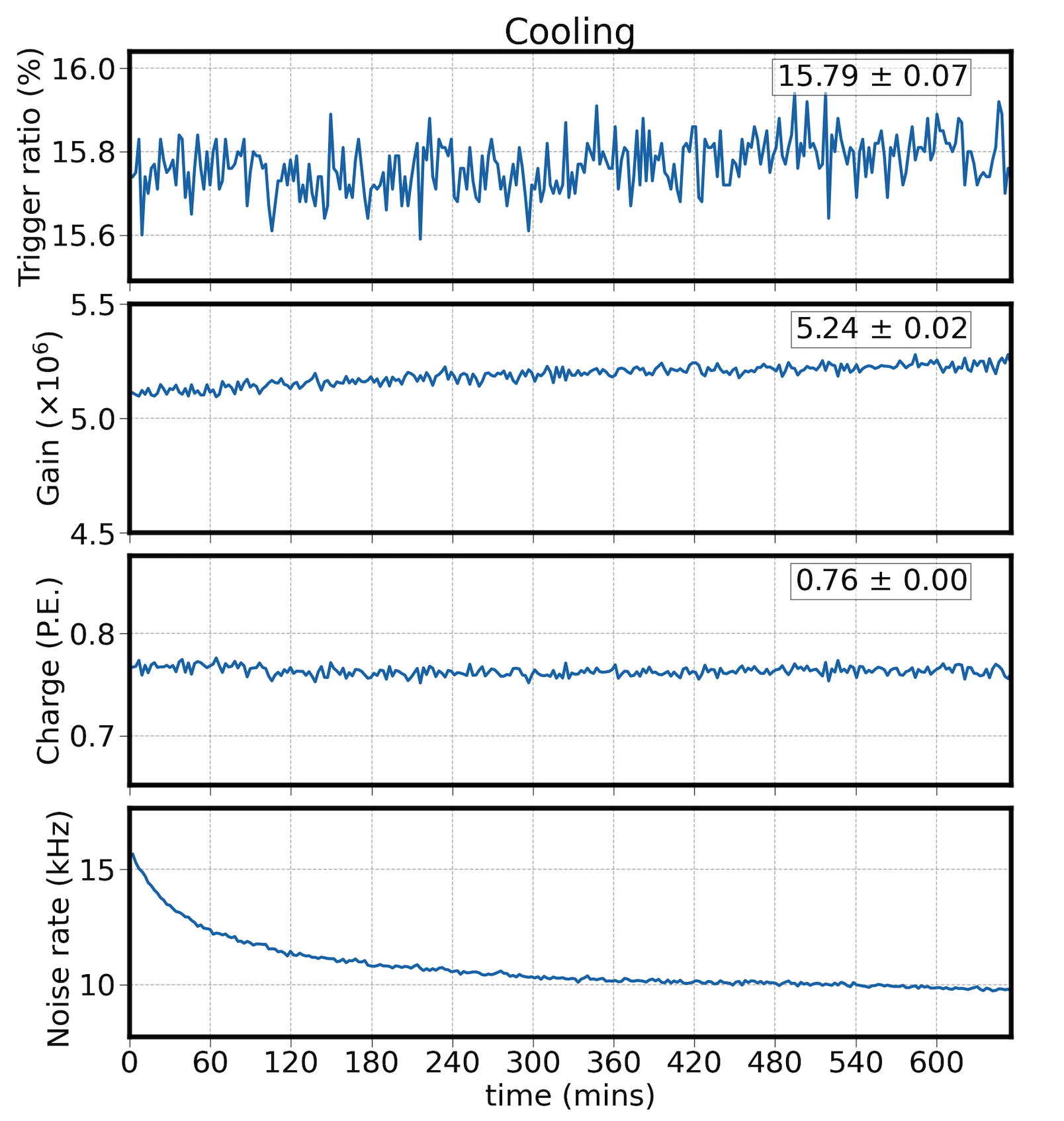}
    \caption{A prolonged cooling test conducted in SPE state for the monitor trigger ratio, gain, charge, and DNR. The mean and fluctuation of the first three are given.}
    \label{fig:cooling}
\end{figure}

1,000,000 trigger signals are collected to measure the trigger ratio, gain, and charge. Subsequently, the LED is turned off, and the TDC is configured for the lowest precision (300 ps/LSB). 5,000,000 trigger windows are measured to determine statistically the signal count under fixed-width conditions. The count is then divided by the effective time to obtain the DNR (with a trigger threshold of 2 mV, 1/3 PE). The above process is repeated, with an execution efficiency of about 140 s/time. In order to ensure that the PMT has sufficient cooling time, based on the results of multiple tests, this needs to be executed at least 50 times, which takes about 2 hours. The final monitoring results are obtained and plotted in figure~\ref{fig:cooling} (using PAB2308-9022 as an example).

The instability during testing comes from the environmental temperature, leakage current, residual gases in the PMT~\cite{affectcooling}. Therefore, consistent working conditions need to be ensured during the testing process. The cooling tests are considered complete when the trigger ratio variation is less than 2\%, gain variation is less than 3\%, charge variation is less than 2\%, and DNR variation is less than 10\%.

\subsection{SPE charge spectrum}
\label{subsec:4.3 SPE charge spectrum}

The SPE charge spectrum test includes measuring four quantities: gain, trigger ratio, SPE resolution, and peak-to-valley ratio. These parameters are tested together because they all need to be measured in the SPE state. The trigger ratio reflects the SPE state of the test, and the SPE resolution and peak-to-valley ratio together represent the PMT’s ability to distinguish single photoelectrons. The switch gear remains in first gear, and the signal is amplified 10 times after being input into N979, synchronously outputting two signals. One signal is still input into V792N, QDC to process the signal.

The setup includes a 20-kHz signal trigger rate, with 2,000,000 trigger signals collected. The V792N is used to record the charge, and from the charge distribution, the trigger ratio, peak-to-valley ratio, and SPE resolution are derived, as shown in figure~\ref{fig:spe_q}. The SPE charge spectrum includes pedestal, gain, SPE resolution, trigger ratio, and peak-to-valley ratio.

\subsection{TTS}
\label{subsec:4.4 TTS}

At the same switch gear as described in Section~\ref{subsec:4.3 SPE charge spectrum}, one signal is output to QDC, while the other signal is input to the leading-edge discriminator N841 and output to V775N. TDC processes the signal.

\begin{figure}[h]
    \begin{minipage}[t]{0.5\linewidth}
        \centering
        \includegraphics[width=\textwidth]{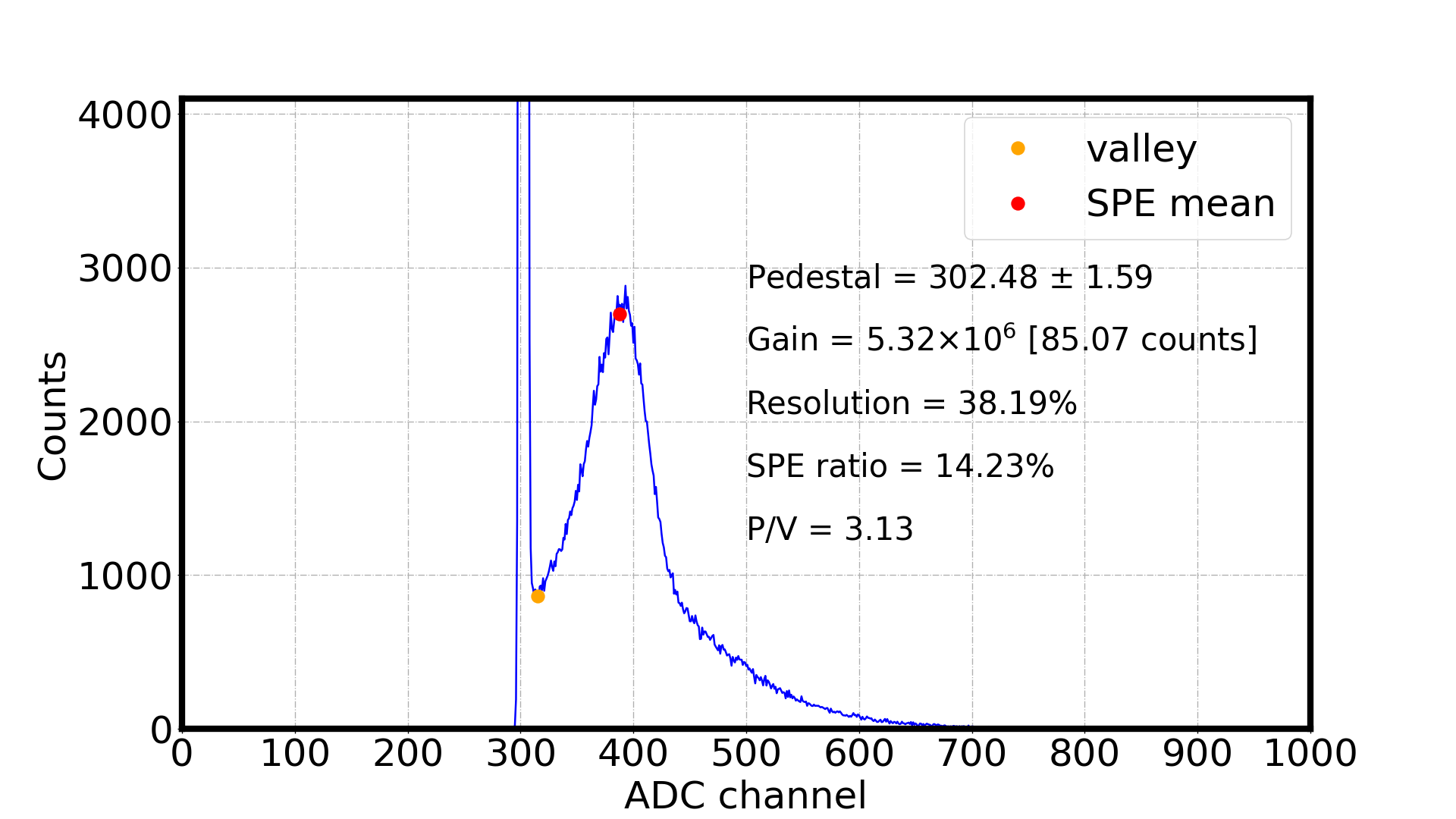}
        \caption{The SPE charge spectrum.}
        \label{fig:spe_q}
    \end{minipage}%
    \begin{minipage}[t]{0.5\linewidth}
        \centering
        \includegraphics[width=\textwidth]{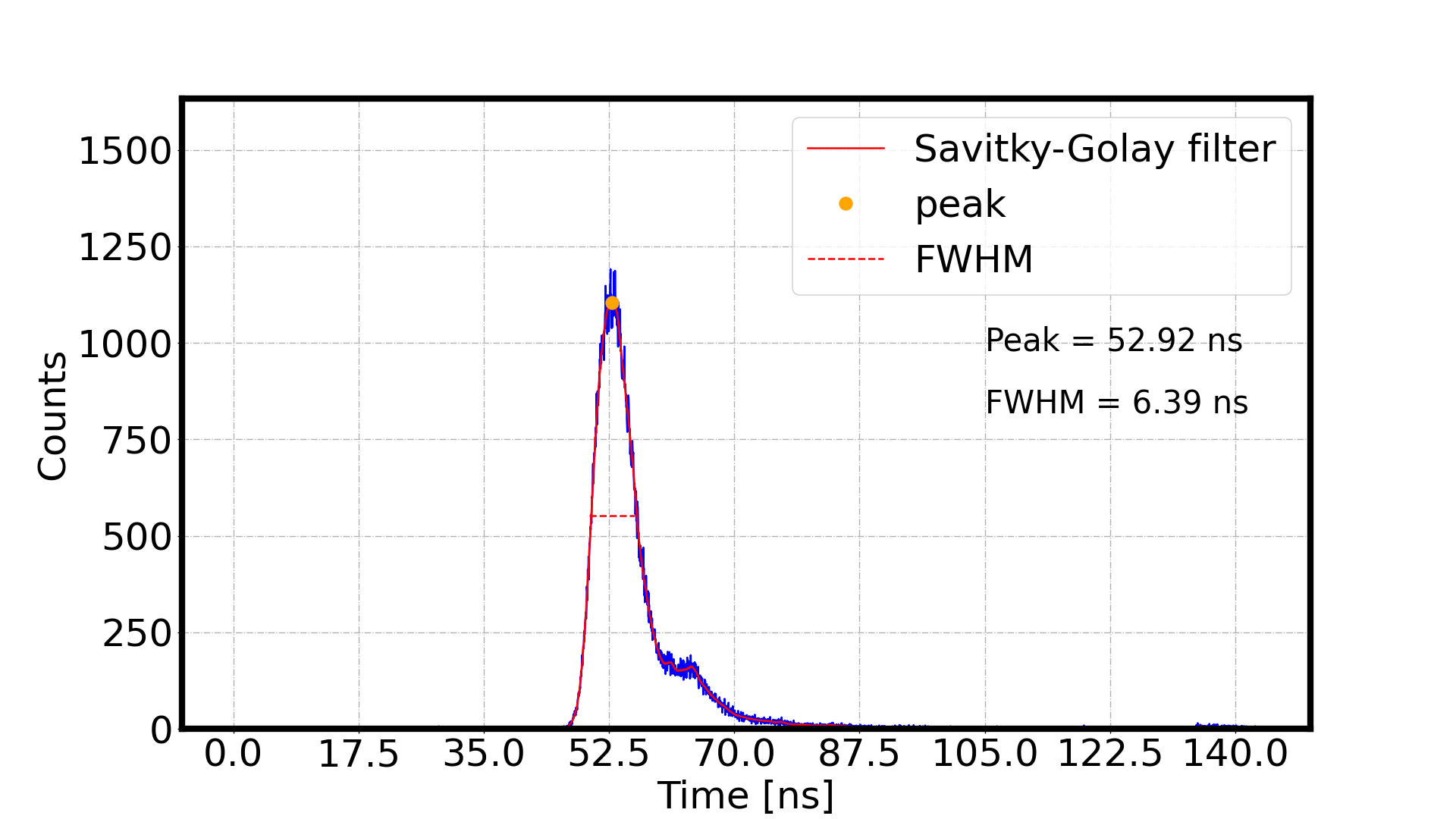}
        \caption{The SPE TTS.}
        \label{fig:spe_t}
    \end{minipage}
\end{figure}

The transition time of the PMT is defined as the time difference between the light reaching the PMT and the signal being received; it follows a Gaussian distribution. The FWHM is defined as the transit time spread, also known as TTS. Due to signal delays caused by signal lines and electronic systems, the directly calculated time difference is commonly represented by relative transit time (RTT). Additionally, the TDC is configured for precision (35 ps/LSB), and 2,000,000 trigger windows are measured to record the TTS. From this distribution, the RTT and TTS are derived, as shown in figure~\ref{fig:spe_t}. After the PMT signal is processed by TDC, the TTS is obtained. The peak position, which is the RTT, is marked in the graph, and the half width, which is the TTS, is also indicated.

\subsection{Q-T effect}
\label{subsec:4.5 q-t effect}

Variation in the signal amplitude from the PMT causes fluctuations in the measurement time, leading to an inconsistency between charge and time, known as the Q-T effect. Therefore, it is necessary to measure the Q-T effect of the PMT and correct the experimental data to eliminate timing errors caused by the leading-edge threshold. The Q-T effect test involves testing both charge and time information. After completing the SPE charge spectrum and TTS test, the command interface of the control computer will prompt to change the switch position to the second setting. Upon completing  this change, the signal is input to the amplifier N978 for a fourfold amplification. The synchronized output is split into two paths. One path enters the V792N, while the other path enters the N841 and then outputs to the V775N. Both the QDC and TDC process the signals.

\begin{figure}[h]
    \centering
    \includegraphics[width=0.7\linewidth]{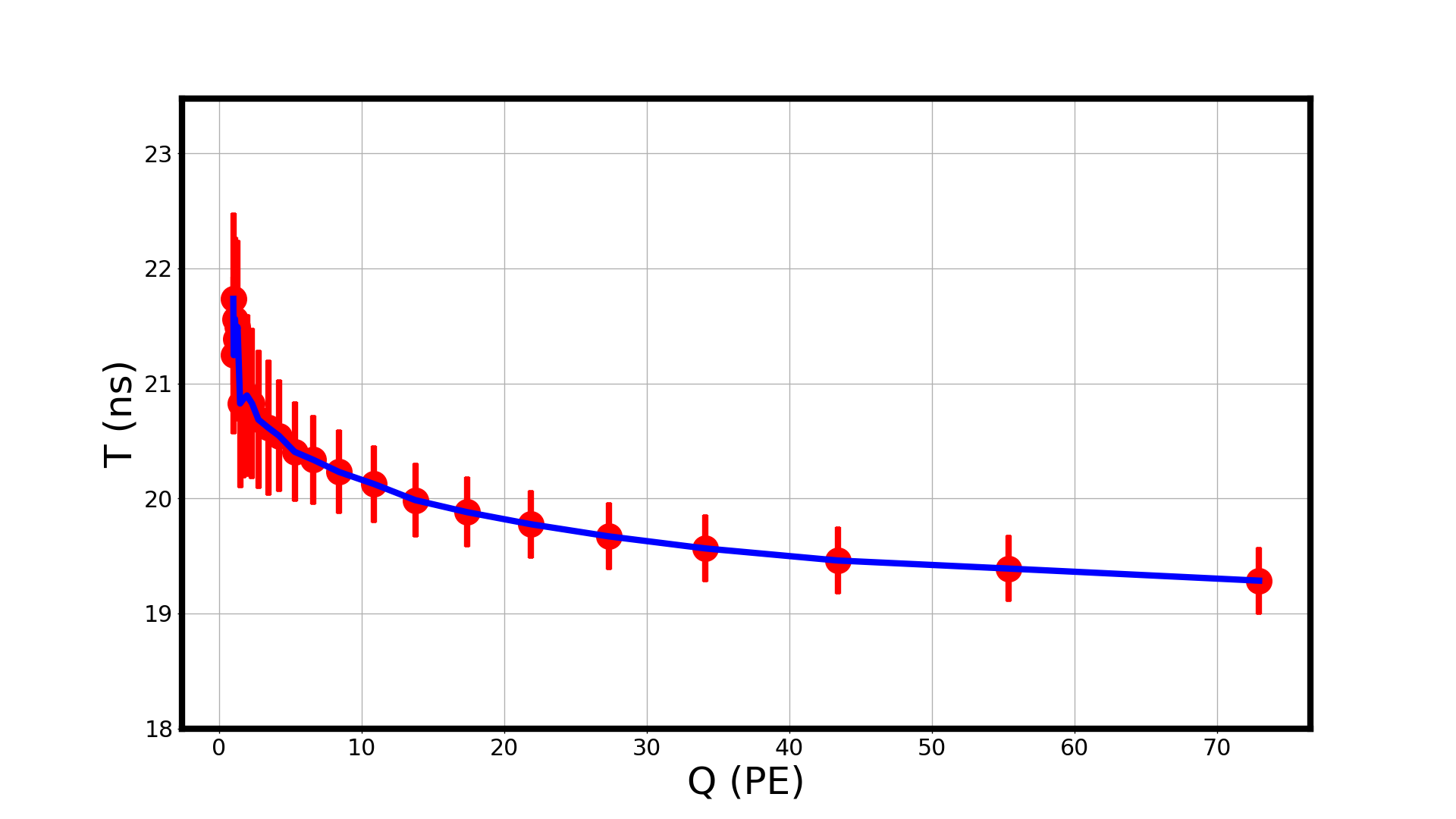}
    \caption{Q-T effect. The relationship between charge and time, with error bars in red.}
    \label{fig:q-t}
\end{figure}

A 2-kHz signal trigger rate is set, and 10,000 trigger signals are collected for both QDC and TDC. The light transmission rate of the attenuator is varied, covering a total of 30 points, and at each point, measurements of both time and charge  are made. The amount of charge and time for each transmittance point are calculated, with a time error of $\text{FWHM}/{\sqrt{12}}$. The FWHM is obtained by fitting the time distribution with a Gaussian. The test results of Q-T effect are shown in figure~\ref{fig:q-t}.

\subsection{Anode charge nonlinearity}
\label{subsec:4.6 anode charge non-linearity}

The anode charge nonlinearity test is primarily aimed at obtaining charge information. Upon completion of the Q-T effect test, the command interface of the control computer prompts to change the switch position to the third setting. After completing the switch operation, the signal is input to the attenuator V859, attenuating the signal by a factor of 10, and then input to the V792N for processing.

At a certain light intensity, the PMT nonlinearity is $L_{att}$, and its output linear charge is $Q_{linearity}$. Under this light intensity, the charge $Q_{test}$ that is obtained is related to the linear charge $Q_{linearity}$ as follows:

\begin{equation}
    Q_{test}=Q_{linearity}\times(1+L_{att})
    \label{con:1}
\end{equation}

\begin{equation}
    Q_{linearity} = I _ {LED} \times E _ { \lambda }\times E_{C}\times G_{HV}\times\tau 
    \label{con:2}
\end{equation}

In the equation \eqref{con:2}, $I_{LED}$ is the light intensity of the LED emission. During testing, the LED emission state remains constant; hence $I_{LED}$ is constant. $\tau$ is the relative light transmittance of the neutral-density filter, which can be obtained from the figure~\ref{fig:led time and light att}. $E_{\lambda}$ is the quantum efficiency for the given wavelength, and $E_{C}$ is the collection efficiency of cathode photoelectrons. $G_{HV}$ is the operating gain of the PMT under the test high voltage. During the test, the high voltage and the peak wavelength of the light source remain constant; thus $E_{\lambda}$, $E_{C}$, and $G_{HV}$ are constants. This means that $Q_{i}$ is positively correlated only with $\tau$.

Typically, at low light intensities, the output of the PMT is close to linear. In that case, the PMT output charge is $Q_{0}$, and the relative transmittance of the attenuator is called $\tau_{0}$. Based on the relative transmittance and the motor steps, the linear charge quantity $Q_{i}$ at any relative transmittance $\tau_{i}$ can be obtained. This can be expressed as

\begin{equation}
    \frac{Q_i}{\tau_i}=\frac{Q_0}{\tau_0}
    \label{con:3}
\end{equation}

The nonlinearity $L_{att}$ at the relative light transmittance $\tau_{i}$ can be obtained from equations \eqref{con:1} and \eqref{con:3}:

\begin{equation}
    L_{att}=\left(\frac{Q_{i}\times\tau_0}{Q_{0}\times\tau_i}-1\right)\times100\%
    \label{con:4}
\end{equation}

\begin{figure}[h]
    \centering
    \includegraphics[width=0.7\linewidth]{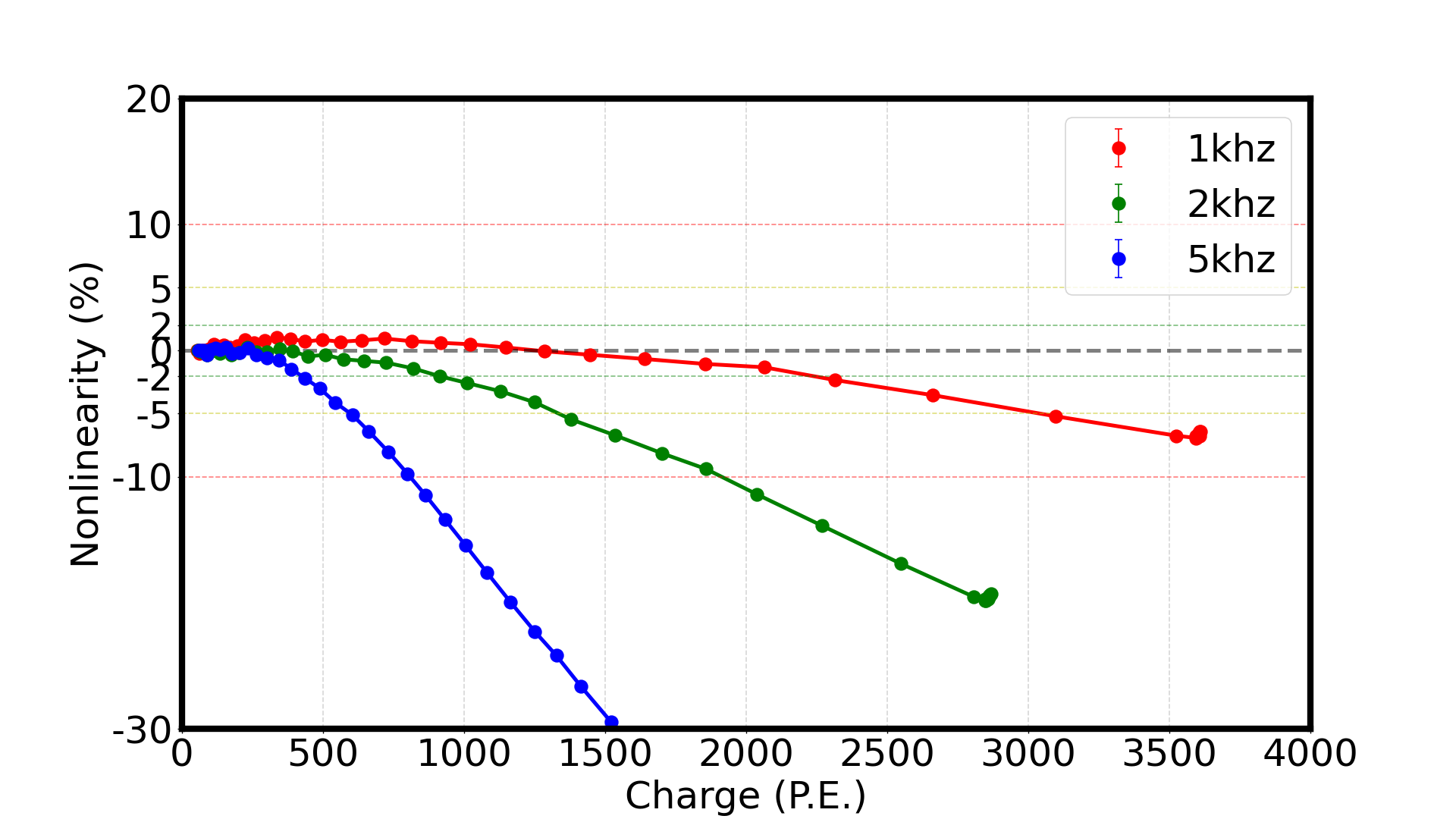}
    \caption{Nonlinear results measured using the optical attenuation method at 1, 2 and 5 kHz of light source.}
    \label{fig:nonl}
\end{figure}

Figure~\ref{fig:nonl} shows the nonlinearity test results obtained using the light attenuation method. The horizontal axis shows the number of photoelectrons nPE calculated based on the actual output charge of the PMT, while the vertical axis shows the nonlinearity calculated using equation \eqref{con:4}. Considering that the charge nonlinearity of the MCP-PMT depends on the signal frequency, three different frequencies (1, 2 and 5 kHz) were used for each measurement to evaluate the nonlinearity.

\subsection{Afterpulse ratio}
\label{subsec:4.7 afterpulse}

\begin{figure}[h]
    \centering
    \includegraphics[width=0.8\linewidth]{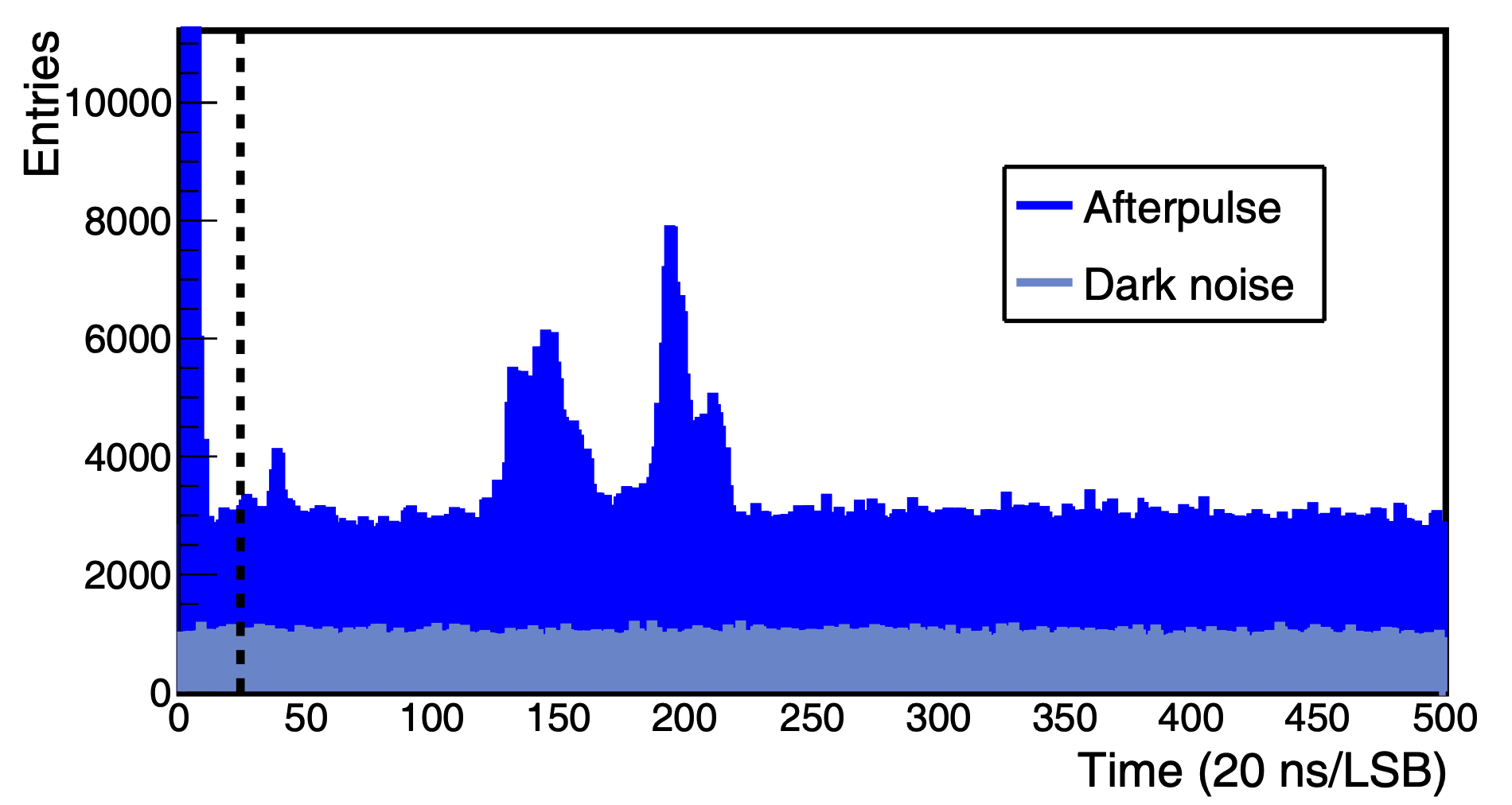}
    \caption{Time distribution of PAB2308-9022 afterpulse. On the right of the black dotted line in the figure is the part of the afterpulse.}
    \label{fig:afterpulse}
\end{figure}

Afterpulse is a signal generated by the ionization of trace gases remaining in the PMT by photoelectrons, which return to the photocathode and are typically microsecond slower than the main pulse~\cite{afterpulse}. The afterpulse ratio test measures the slow afterpulse components in the PMT signal. According to Section~\ref{subsec:4.1 waveform check}, SPE corresponds to 6 mV, and the signal amplitude displayed by the PMT on the oscilloscope is calibrated to 180 mV, that is, 30 PE, by adjusting the light intensity through an optical attenuator. This is because in the SPE state, the PMT signal-to-noise ratio of PMT is relatively low, resulting in significant test errors. According to the conclusion in Section~\ref{subsec:4.6 anode charge non-linearity}, the level of 30 PEs does not cause any non-linearity in the test results. The PMT signal is then input into the N979 and amplified tenfold. It is next input into the CIQTEK TDC0110. This TDC module can adjust the threshold through PC driver software, with the threshold set to 1/3 PE. The plotting window is set to 10 $\mu$s, and the acquisition time is set to 4 minutes. Clicking the start button to begins the test; the results are shown in figure~\ref{fig:afterpulse}. 

The afterpulse ratio is calculated using equation \eqref{con:5}. The count from 0.5 $\mu$s to 10 $\mu$s after the main pulse is determined statistically and called $N_A$. Under the same conditions, DNR measured during the same time window when the light source is turned off is $D_R$ $\times$ 9500 $\times$ 10$^{-9}$. $N_0$ is the count of the main pulse. The denominator of 30 is used to normalize the result to SPE. The afterpulse ratio given by equation \eqref{con:5}, taking PAB2308-9022 as an example, is 0.086\% $\pm$ 0.001\%.

\begin{equation}
    \text{Ratio}=\frac{N_A-D_R\times9500\times10^{-9}}{30\times N_0}\times100\%
    \label{con:5}
\end{equation}

\section{Summary}
\label{sec:5}

\begin{figure}[ht]
    \centering
    \includegraphics[width=0.9\linewidth]{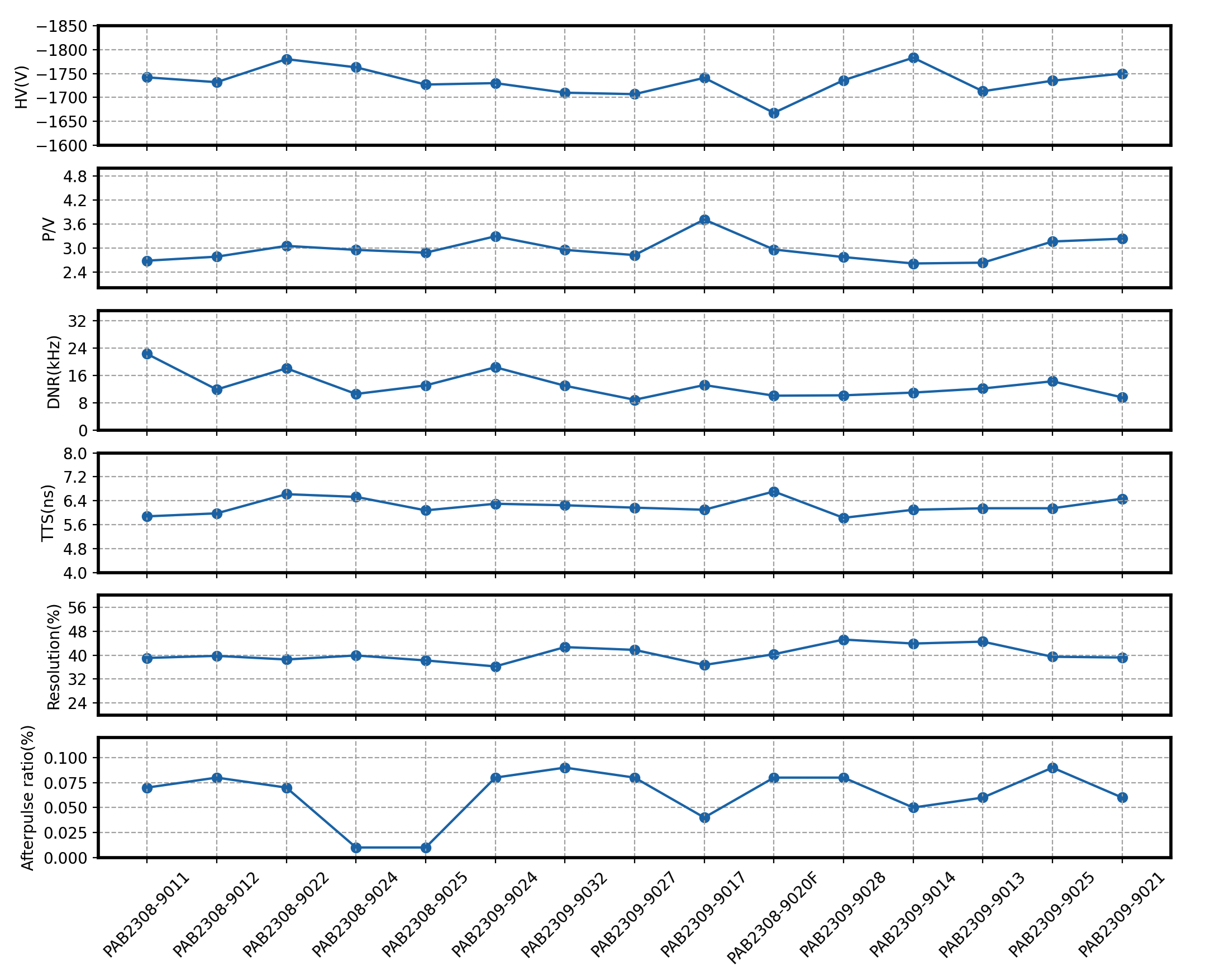}
    \caption{The test results of HV, peak-to-valley ratio, DNR, TTS, and SPE resolution and Afterpulse ratio.}
    \label{fig:distribution}
\end{figure}

We set up a set of PMT test platform and established the test process of PMT performance. The test results of HV, peak-to-valley ratio, DNR, TTS, and SPE resolution and Afterpulse ratio are summrized in figure~\ref{fig:distribution}. These parameters will be used to evaluate the performance of PMTs for the HUNT experiment. All PMTs operate at a gain of 5 $\times$ 10$^6$; the high voltages range vary from 1650 to 1800 V, peak-to-valley ratios all are larger than 2.0, and SPE resolutions are smaller than 50\%. After a certain cooling period, the DNRs become below 25 kHz. And TTS are less than 7 ns. Afterpulse ratios of the tested PMT are all below 0.1\%. Table~\ref{tab:summary} gives the statistical analysis of the 15 PMT test results mentioned above. Therefore, all above test results have met the requirements.

\begin{table}[ht]
\centering
\caption{The mean and standard deviation of the measured 15 PMTs.}
\smallskip
\begin{tabular}{l l l} 
  \hline
  Parameter & Value \\ 
  \hline
  HV (V) & -1743 $\pm$ 28 \\ 
  Peak-to-valley & 2.96 $\pm$ 0.28 \\ 
  DNR (kHz) & 13.13 $\pm$ 3.65 \\ 
  TTS (ns) & 6.22 $\pm$ 0.25 \\ 
  SPE Resolution (\%) & 40.37 $\pm$ 2.61 \\ 
  Afterpulse ratio (\%) & 0.06 $\pm$ 0.02 \\ 
  
  \hline
\end{tabular}
\label{tab:summary}
\end{table}

This article mainly focuses on testing the performance of a new type of 20-inch PMT developed for the HUNT experiment, in preparation for the prototype deployment plan in March 2024. Besides, the testing system established in this work will provide the foundation for testing a larger batch of PMTs for the HUNT experiment in the future. From the first batch of PMTs delivered by NNVT, 15 PMTs with qualified performance have been selected and are awaiting assembly into the OMs.


\acknowledgments

This work is supported by the Science and Technology Department of Sichuan Province (Grant No. 2021YFSY0030), and the IHEP funds (Grant No. E42982U8 \& E35452U2).




\bibliographystyle{JHEP} 
\bibliography{refs.bib}

\end{document}